\documentclass[showpacs,floatfix,twocolumn]{revtex4-1}
\usepackage{graphicx,psfrag,amsmath,amssymb,amsfonts,latexsym,color,epsf,graphpap,dcolumn}
\begin{document}
\title{On the Difference Between the Vacuum Casimir Energies for Grounded
and Isolated Conductors}
\author{C.~D.~Fosco$^{a}$, F.~C.~Lombardo$^{b}$ and F.~D.~Mazzitelli$^{a}$}
\affiliation{$^a$Centro At\'omico Bariloche and Instituto Balseiro, CONICET,
Comisi\'on Nacional de Energ\'\i a At\'omica, R8402AGP Bariloche, Argentina.\\
$^b$Departamento de F\'\i sica {\it Juan Jos\'e
Giambiagi}, FCEyN UBA and IFIBA CONICET-UBA, Facultad de Ciencias Exactas y Naturales,
Ciudad Universitaria, Pabell\' on I, 1428 Buenos Aires, Argentina}
\date{today}
\begin{abstract}
\noindent 
We study the vacuum (i.e., zero-temperature) Casimir energy for a system of
neutral conductors which are {\em isolated\/}, as opposed to {\em
grounded}. The former is meant to describe a situation where the total
charge on each conductor, as well as all of its fluctuations, vanishes,
while the latter describes a situation where the conductors are connected
to a charge reservoir.   We compute the difference between the vacuum
energies for a given system of conductors, but subjected to the two
different conditions stated above. The results can be written in terms of a
generalized, frequency-dependent capacitance matrix of the system. Using a multipolar 
expansion, we show that the grounded Casimir energy includes a monopole-monopole interaction term that is absent in the isolated case in the large distance limit. 
\end{abstract} 
\pacs{12.20.Ds, 03.70.+k, 11.10.-z}
\maketitle
\section{Introduction}\label{sec:intro}

In this article we consider the  Casimir energy  for the quantum
electromagnetic (EM) field in the presence of a system of perfect
conductors \cite{books},  paying particular attention to the  boundary conditions
allowed by perfect conductivity.  At the classical level, perfect
conductivity does not specify the problem completely, unless one
introduces some extra `global' information regarding the state of those conductors.
For example, in the electrostatic case, it is necessary to specify either
the total charge or the value of the electrostatic potential on each conductor. In time
dependent situations one should specify the charge of each conductor, or
the values of the EM four-potentials, once the gauge freedom is properly
fixed.  At the quantum level, that extra  information will play the role of
`boundary conditions' for the fluctuating fields.  Among the possible
choices for the state of the conductors, we will focus on two cases:
grounded and isolated (and globally neutral) conductors.

In the functional integral approach to the calculation of the Casimir
energy,  the fluctuating EM  field is integrated out subjected to the
proper gauge and boundary conditions, which for perfect conductors imply
that the  tangential component of the electric field and the normal
component of the magnetic field vanish on the surfaces of the conductors.
These conditions alone do not fix completely the total charge of each
conductor: although the mean value of the charges vanish, the fluctuations
do not. Physically, this situation corresponds to a case in which the
conductors are connected to a charge reservoir (we will refer to this
situation as grounded). One could also consider the alternative  case in
which the conductors are neutral and isolated. This means that not only the
mean value of the total charge should vanish, but also do all correlations
between total charges. Mathematically, it is necessary to introduce
additional constraints in the path integral, to insure that the total
charges do not fluctuate. Therefore, the Casimir  energy will be different 
for grounded and isolated conductors. The main goal of this paper will be
to obtain an explicit expression for the difference between grounded and
isolated Casimir  vacuum energies at zero temperature.  
This problem has been partially addressed in previous works
\cite{Golestanian2000, Fosco:2016rpu}. In particular, in
Ref.\cite{Fosco:2016rpu} we considered the Casimir free energies in the
high temperature limit.  In that case, only the zero Matsubara  frequency
of the EM field is relevant and the problem can be described in terms of an
``electrostatic potential". For grounded conductors, the potential vanish
on each surface, and the total charge of each conductor is not fixed. On
the other hand, for isolated bodies, the total charges vanish and do not
fluctuate, while the values of the potential on each surface are not
specified. Moreover,  the partition function for the isolated case can be
obtained from a partition function in which one specifies the values of the
potential on each conductor, and then integrates out over all those
possible values.  
 
 In order to obtain the partition function for grounded and isolated
 conductors for the EM field at zero temperatures, in   
Section~\ref{sec:iso}  we will  first revisit the high temperature limit, and show that the partition function for the isolated case can be obtained by imposing both perfect conductor boundary conditions on the surfaces along with the additional constraints of vanishing total charges on each conductor. Indeed, in the grounded case the conductors are connected to charge reservoirs, and therefore one should include in the functional integral  field configurations with arbitrary values of the total charges. On the contrary, in the isolated case,  the functional integral is restricted  only to those configurations with vanishing total charge in each conductor. As we will see, this procedure can be implemented {\it mutatis-mutandis} to the EM field at zero temperature. 

 In Section~\ref{sec:thesys}, we introduce the observable that we study
in the remainder of this article, namely, the difference between the
Casimir energies for grounded or isolated conductors at $T=0$. We find an explicit expression for this difference in terms of generalized, frequency-dependent capacitance coefficients.  

In Section~\ref{sec:res} we evaluate the difference of the Casimir energies in the large distance limit, using a multipolar expansion. As expected,
the grounded Casimir energy includes a monopole-monopole interaction that is absent in the isolated case, and therefore it shows a slower decay with distance.

We present our conclusions in Section V. In particular, we discuss the relation between our results and  those obtained
for perfect conductors as a limiting case of magnetodielectric materials.  

\section{Grounded versus isolated conductors: partition functions}\label{sec:iso}
\subsection{Electrostatics}
In order to clarify the further developments, we begin by considering a
simplified version of the system that we want to consider, namely, that of
electrostatics in the presence of a system of conductors. Fluctuations are
introduced into the system by regarding it as in thermal equilibrium with a heat bath at a
given temperature, and working with the corresponding canonical ensemble. Note
that this classical system may also be approached by taking the high-temperature
limit for the finite-temperature quantum EM field version. This is indeed the case
we have considered, from a different standpoint, in a previous
reference~\cite{Fosco:2016rpu}.

To introduce the canonical ensemble, we need its classical energy $E$,
which we will write in terms of the electrostatic potential.
To define the geometry of the system, we denote by $V_\alpha$
($\alpha = 1,\dots,N$) the spatial region
occupied by each conductor and by $S_\alpha$ the boundary (a
closed surface) of each region, respectively.  
Besides, we use $U$ for the complement of the region occupied by
the conductors; namely, \mbox{$U = {\mathbb R}^3 - \cup_{\alpha=1}^N
V_\alpha$}. If $S$ denotes the boundary of $U$, then 
\mbox{$S= \cup_{\alpha=1}^N S_\alpha$}. 

For  electrostatics the energy functional reads:
\begin{equation}\label{eq:defephi}
	E[\phi] \;=\; \int_U d^3x \,\frac{1}{2} \, | \nabla
	\phi({\mathbf x})|^2 \;,
\end{equation}
in terms of the electrostatic potential $\phi$.

The classical partition function for isolated and grounded conductors can be written
in terms of an intermediate object, which is  a
partition function where the (constant) value of $\phi$ on each surface $S_\alpha$ is
fixed to a given but otherwise arbitrary value $\phi_\alpha$:
\begin{equation}
 {\mathcal Z}[\{\phi_\alpha\}]=\, \int {\mathcal D} \phi \,
e^{- \beta E[\phi]}
\prod_{\alpha=1}^N\delta[\phi\vert_{S_\alpha}-\phi_\alpha] \; ,
\end{equation}
where $\beta = \frac{1}{T}$ (note that in Ref.\cite{Fosco:2016rpu} we used a different normalization for the electrostatic potential). From now on,
we set $\beta=1$,  since the global factor in the action will be irrelevant for our discussion.

Thus, we may obtain the partition functions corresponding to  grounded
(${\mathcal Z}^{(g)}$) and isolated (${\mathcal Z}^{(i)}$) conductors as follows \cite{Golestanian2000,Fosco:2016rpu}:
\begin{equation}
{\mathcal Z}^{(g)}\;=\;{\mathcal Z}[\{\phi_\alpha \}]\Big|_{\phi_\alpha =
0}
\end{equation}
and
\begin{equation} \label{iso}
{\mathcal
Z}^{(i)}\;=\;\int_{-\infty}^{\infty}\left(\prod_{\alpha=1}^{N}d\phi_\alpha\right)
\; {\mathcal Z}[\{\phi_\alpha \}]\;.
\end{equation}

The partition function ${\mathcal Z}^{(i)}$ can be derived using an alternative procedure, that will be useful for the  generalization to 
the case of electrodynamics.  We will impose, in addition to the Dirichlet boundary conditions, the constraint
of vanishing total charge on each conductor
\begin{equation}
{\mathcal Z}^{(i)}\;=\;  \int {\mathcal D} \phi \,
e^{-  E[\phi]}
\prod_{\alpha=1}^N\delta[\phi\vert_{S_\alpha}]\delta[Q_\alpha]\; , 
\label{Zialt}
\end{equation}
where 
\begin{equation} \label{qelec}
Q_\alpha=-\oint_{S_\alpha}d s_\alpha\, \nabla\phi\cdot {\mathbf n_\alpha}\, \,\,\, ,
\end{equation}
 and ${\mathbf  n_\alpha}$ is the unit outer normal to the surface.
Strictly speaking, the integral in Eq.(\ref{qelec}) should be performed on a surface that is outside and infinitesimally close to the conductor's surface, since the normal component of the electric field is discontinuous across the surface. This detail will be implicitly understood in what follows. The $\delta$-function on each surface can be written in terms of $N$ auxiliary variables $\phi_\alpha$ as:
\begin{equation}
2 \pi\,  \delta[Q_\alpha]=\int  d\phi_ \alpha  \, e^{i\phi_\alpha Q_\alpha}=\int    d\phi_ \alpha \, e^{i\int d^3{\mathbf x}\, \rho_\alpha\, \phi}\, ,
\end{equation}
where
 \begin{equation}
 \rho_\alpha=-\phi_\alpha\oint_{S_\alpha}ds_\alpha\,  {\mathbf  n_\alpha}\cdot\nabla\delta({\mathbf x}-{\mathbf x_{S_\alpha}})\, \,  .
 \end{equation}
 Note that $\rho_\alpha$ corresponds to the surface charge density of an electric dipole layer of strength $\phi_\alpha{\mathbf n}_\alpha$  which is infinitesimally close to the surface of the conductor. As discussed previously when studying the effect of patch potentials on the Casimir force \cite{patches}, the combination of a grounded conductor and  such a layer is equivalent to a single conductor at a potential $\phi_\alpha$. Therefore,  the two constraints in Eq.(\ref{Zialt}) are equivalent to impose first $\phi\vert_{S_\alpha}=\phi_\alpha$ and then integrate over all possible values of $\phi_\alpha$. 
 Thus, Eqs.(\ref{iso}) and (\ref{Zialt}) are equivalent. One can check this explicitly by performing the functional integral in Eq.(\ref{Zialt}). The result is
 \begin{equation}
 {\mathcal Z}^{(i)} =  \Big[
\prod_{\alpha=1}^N\int d\phi_\alpha e^{\frac{1}{2}\int d^3{\mathbf x}\int d^3{\mathbf y}\sum_{\alpha\beta}\rho_\alpha({\mathbf x})G({\mathbf x},{\mathbf y})\rho_\beta({\mathbf y})}\Big] {\mathcal Z}^{(g)} 
 \end{equation}
 where $G$ is the Green's function of the electrostatic problem 
satisfying Dirichlet boundary conditions.  
 Using the explicit expression for $\rho_\alpha$, and after integration by parts we obtain
 \begin{eqnarray}\label{JGJ}
 &&\int d^3{\mathbf x}\int d^3{\mathbf y}\rho_\alpha({\mathbf x})G({\mathbf x},{\mathbf y})\rho_\beta({\mathbf y}) \\
&& = \phi_\alpha\phi_\beta \oint _{S_\alpha}ds_\alpha \oint_{S_\beta} ds_\beta\;
 \partial_{n_\alpha} \partial_{n_\beta}G 
 = -  \phi_\alpha\phi_\beta  C_{\alpha \beta}  , \nonumber
  \end{eqnarray} 
 where $C_{\alpha\beta}$ are the coefficients of capacitance (written in terms of the Green's function). Therefore
 \begin{equation}
 {\mathcal Z}^{(i)}=\Big[
\prod_{\alpha=1}^N\int d\phi_\alpha\, e^{-\frac{1}{2}\sum_{\gamma\delta}C_{\gamma\delta}\phi_\gamma\phi_\delta}\Big]
\, {\mathcal Z}^{(g)}\; ,
\end{equation} 
which is the same result we obtained in our previous paper \cite{Fosco:2016rpu},  starting from Eq.(\ref{iso}).

\subsection{Electrodynamics} 

We deal now with the equivalent situation but in electrodynamics at zero
temperature. 
In our conventions, the Euclidean Maxwell action ${\mathcal
S}$  is given by:
\begin{equation}\label{eq:defsa}
	{\mathcal S} \;=\; - \frac{1}{4}\int dx_0\int_U d^3{\mathbf x}\, \,F_{\mu\nu} F_{\mu\nu} 
\end{equation}
with $F_{\mu\nu} \equiv \partial_\mu A_\nu - \partial_\nu A_\mu$.  The space-time indices $\mu,
\nu,\ldots$ run from $0$ to $3$; the Euclidean time coordinate is
$\tau=x_0$.

Due to gauge invariance,  the generalization of Eq.(\ref{iso}) to the EM field  is not obvious. On the contrary, Eq.(\ref{Zialt})
admits such generalization. Indeed, let us define the grounded (Euclidean) partition function as
\begin{equation}
	{\mathcal Z}^{(g)} \;=\; \int {\mathcal D}A \; \delta_G[A]
\;	e^{-{\mathcal S}[A]}
\end{equation}
where ${\mathcal S}$ is the Maxwell action and $\delta_G[A]$ is a shorthand notation for the
perfect-conductor boundary conditions imposed on the conductors
(gauge fixing is implicitly understood in the integration measure).

For an isolated system, on the other hand, one has to introduce a
constraint on the total charge in each conductor.  This can be introduced
on top of the previous case by adding to the integration measure a product
of $\delta$-functions of $Q_\alpha$:
\begin{equation}\label{isoEM}
	{\mathcal Z}^{(i)} \;=\; \int {\mathcal D}A \,\delta_G[A]\delta_Q[A]
	\, e^{- {\mathcal S}[A]} \; ,
\end{equation}
where
\begin{equation}
\delta_Q[A]=  \prod_{\alpha=1}^N \delta( \oint d{\mathbf s}_\alpha \cdot
{\mathbf E}_{(\alpha)} ) \, .
\end{equation}

Note that we can write Eq.(\ref{isoEM}) as:
\begin{equation}\label{eq:zi1}
{\mathcal Z}^{(i)} \;=\; {\mathcal Z}^{(g)} \; \left\langle  \, \delta_Q[A]
 \right\rangle_G 
\end{equation}
where the `$G$-average' symbol is defined by:
\begin{equation}
\big\langle \ldots \big\rangle_G \;\equiv\; 
\frac{\int \big[{\mathcal D} A\big] \;\delta_G[A] \;\ldots \; e^{-
{\mathcal S}[A]}}{\int \big[{\mathcal D} A\big] \;\delta_G[A] \;
e^{-{\mathcal S}[A]}} \;.
\end{equation}
Eq.(\ref{eq:zi1}) will be our starting point to analyze the difference between the zero temperature Casimir energies for grounded and isolated
conductors

\section{Grounded versus isolated Vacuum energies}\label{sec:thesys}

Let us begin by introducing $E^{(g)} (E^{(i)})$, the vacuum energy for the EM field
in the presence of a system of grounded (isolated) conductors. We may write that
energy in terms of the corresponding zero-temperature  Euclidean
partition function ${\mathcal Z}^{(g)}({\mathcal Z}^{(g)})$ , as follows:
\begin{equation}
E^{(g,i)} \;=\; - \, \lim_{T \to \infty} \big[ \frac{1}{T} \log {\mathcal Z}^{(g,i)} \big]
\;,
\end{equation}
where $T$ is the extent of the Euclidean time coordinate. Therefore, the observable we are interested in is
\begin{equation}
\Delta E \equiv E^{(g)} - E^{(i)}\;=\; \lim_{T \to \infty} \Big[ \frac{1}{T} \log \left\langle  \, \delta_Q[A]
\, \right\rangle_G \Big] \;.
\end{equation}

To proceed, it is necessary to be more explicit about the boundary conditions on each surface. In particular, 
we need an explicit functional expression for $\delta_Q[A]$. Let us
consider one of the surfaces, and introduce a
parametrization for its points,  in terms of two real parameters
$(\sigma^1, \sigma^2)\equiv \sigma$: 
\begin{equation}
\sigma \; \to \; {\mathbf y}(\sigma) \;,
\end{equation}
where ${\mathbf y} \in {\mathbb R}^3$.  Let us also introduce the notations $y^\mu=(x_0,{\mathbf y})$ and $\sigma^a=(\sigma^0\equiv x_0,\sigma^1,\sigma^2)$. Latin indices $a,b,c$ run from $0$ to $2$. 

Perfect conductor boundary conditions can be written in terms of the projection of the vector potential on the surface as
\begin{equation}\label{pcgen}
	\epsilon_{abc} \partial_b {\mathcal A}_c \vert_S\;=\; 0 \, ,
\end{equation}
where $ {\mathcal A}_c=e^{c}_\mu A_\mu$  and
$e^{c}_\mu = \partial y^\mu/\partial\sigma^c$. 
Then, the vanishing of ${\mathbf
E}_\parallel$, the components of the electric field which are parallel to
the surface, is written as:
\begin{equation}
\frac{\partial {\mathbf y}}{\partial \sigma^i} \cdot {\mathbf
E}(\tau, {\mathbf y}(\sigma) ) \;=\;0 \;\;,\;\;\;\;  i \,=\, 1, 2 \;,
\end{equation}
(${\mathbf E} \equiv \partial_0 {\mathbf A} - \nabla A_0$), while for the
normal component of the magnetic field ${\mathbf B} \equiv \nabla \times
{\mathbf A}$, the corresponding condition is:
\begin{equation}
{\mathbf n}(\sigma) \cdot {\mathbf B}(\tau, {\mathbf y}(\sigma) )
\;=\;0 \;\;,
\end{equation}
with ${\mathbf n}(\sigma)$ the unit normal to $S$ at the point ${\mathbf
y}(\sigma)$: 
\begin{equation}
{\mathbf n}(\sigma) \;=\; \frac{{\mathbf N}(\sigma)}{|{\mathbf N}(\sigma)|} 
\;,\;\;
{\mathbf N}(\sigma) \;=\; \frac{\partial {\mathbf y}}{\partial
\sigma^1} \times \frac{\partial {\mathbf y}}{\partial \sigma^2} \;.
\end{equation}

For the isolated partition function, we select, among the field
configurations with vanishing parallel components of the electric field and
normal components of the magnetic field on the surfaces of the conductors,
only those where the total charge on each one them vanishes. Contrary
to what happens for the perfect conductor boundary conditions,
this condition is nonlocal, since it involves
an integral over each surface:
\begin{equation}
	Q_\alpha(\tau) \;=\; \int_{S_\alpha} \, d^2 \sigma \, {\mathbf
	N}(\sigma) \cdot {\mathbf E}(\tau,{\mathbf y}(\sigma)) \;=\; 0 \;. 
\end{equation}

It is convenient to introduce $N$ auxiliary time-dependent
fields $\lambda_\alpha(\tau)$ ($\alpha = 1, \ldots, N$), so that:
\begin{equation}
	\delta_Q[A] \;=\; \int \big(\prod_{\alpha=1}^N \, {\mathcal D}
	\lambda_\alpha \big) \; e^{ - i \int_{-\infty}^{+\infty} 
	d\tau \, \sum_{\alpha=1}^N \lambda_\alpha(\tau) \, Q_\alpha(\tau)}
	\;.
\end{equation}
By an integration by parts, we see that 
\begin{equation}
	\delta_Q[A] \;=\; \int \big(\prod_{\alpha=1}^N \, {\mathcal D}
	\lambda_\alpha \big) \; e^{  i \, \int d^4x J_\mu(x) \, A_\mu(x) } \;,
\end{equation}
with
\begin{eqnarray}\label{eq:defj}
	J_0(x) &=& \sum_{\alpha=1}^N \lambda_{\alpha}(\tau) \, \sum_i
	N_i^\alpha(\sigma) \,\frac{\partial}{\partial x_i} \delta({\mathbf
	x} - {\mathbf y}^\alpha(\sigma)) \nonumber\\
	J_i(x) &=& - \sum_{\alpha=1}^N \dot{\lambda}_{\alpha}(\tau) \, 
	N_i^\alpha(\sigma)
	\, \delta({\mathbf x} - {\mathbf y}^\alpha(\sigma)) \;,
\end{eqnarray}
where we have added an $\alpha$ index to distinguish objects belonging to
different surfaces.

Now, recalling that the integral over the gauge field in the $\langle
\ldots \rangle_G$ averages is a Gaussian, we see that $\Delta E$ may be
represented as follows:
\begin{equation}
	\Delta E \;=\; \lim_{T \to \infty} \log \big(\frac{1}{T} \Delta
	{\mathcal Z} \big) \;,
\end{equation}
where
\begin{equation}
	\Delta{\mathcal Z}\;=\;\int \big(\prod_{\alpha=1}^N \, {\mathcal D}
	\lambda_\alpha \big) \, e^{-\frac{1}{2} \int d^4x \int d^4x'
	J_\mu(x) G_{\mu\nu}(x,x') J_\nu(x')} \;,
\end{equation}
where $J_\mu$ is as defined in (\ref{eq:defj}) and we have introduced
$G_{\mu\nu}$, the $\langle A_\mu(x) A_\nu(x')\rangle_G$ correlation function
for the gauge field in the presence of the grounded conductors.

Taking into account the specific form of $J_\mu$, we see that:
\begin{equation}\label{deltaEfin}
	\Delta E \;=\; - \frac{1}{2} \, \int \frac{d \omega}{2\pi} \, \log[
	\det \widetilde{\mathbb C}(\omega) ] \;,
\end{equation}
where we have introduced a frequency-dependent matrix ${\mathbb C}(\omega)
= [C_{\alpha\beta}(\omega)]_{N\times N}$, which generalize the capacitance
coefficients:
\begin{equation}
	\tilde{C}_{\alpha\beta}(\omega) = \int d\tau e^{i \omega \tau}
	C_{\alpha\beta}(\tau) \, ,
	\end{equation}
with
\begin{equation}	
	C_{\alpha\beta}(\tau) = \sum_{i,j} \oint ds_i^\alpha \oint ds_j^\beta \;
	\langle E_i(\tau,{\mathbf y}^\alpha)  E_j(0,{\mathbf y}^\beta)
	\rangle_G \; .
\end{equation}
Here $ds_i^\alpha$ denotes $i$-component of the outer normal area element for the conductor
surface labeled by $\alpha$. Eq.(\ref{deltaEfin}) is our main result.

Recalling the relation between the normal component of the electric field
and the surface density in a conductor, we may also write:
\begin{equation}
	C_{\alpha\beta}(\tau) \;= \; \oint ds^\alpha \oint ds^\beta \;
	\langle \sigma(\tau,{\mathbf y}^\alpha)  \sigma(0,{\mathbf y}^\beta)
	\rangle_G \;.
\end{equation}
Note that the time dependent capacitance
coefficients introduced here reduce to the usual ones in the electrostatic case (see Eq.(\ref{JGJ})). 
Therefore,  the formulae above reproduce the high temperature ones
in the case of periodic time evolution, if one just keeps the zero mode.

\section{Multipolar expansion of the grounded vacuum energy}\label{sec:res}

In order to illustrate the main difference between the grounded and isolated vacuum energies, in this section we will compute $E^{(g)}$ using a multipolar expansion, which is adequate to describe the interaction between conductors separated at distances much larger than their characteristic
sizes.   The main contribution in this expansion is the monopole-monopole interaction, which is absent for isolated conductors due to the condition of neutrality.
All the other contribution, coming from higher multipoles, are equal for
grounded or isolated conductors, since they are independent of the total
charge (we assume multipoles are defined with the center of charge as the
origin).

To perform the calculation, it is necessary to obtain an explicit form for the constraint $\delta_G[A]$. Taking into account
Eq.(\ref{pcgen}), and introducing an auxiliary field
$\xi_a^{(\alpha)}(x_0,\sigma)$ on each surface we have

\begin{eqnarray}
\delta_G[A] &=& \int\Big[\prod_{\alpha=1}^N{\mathcal D}\xi^{(\alpha)}_a\Big]e^{i\sum_\alpha\oint_{S_\alpha}\xi_a^{(\alpha)}\epsilon_{abc} \partial_b {\mathcal A}^{(\alpha)}_c} \nonumber \\
&\equiv &  \int\Big[\prod_{\alpha=1}^N{\mathcal D}\xi^{(\alpha)}_a\Big]e^{i\int d^4x \, J_\mu A_\mu}\, , \end{eqnarray}
where 
\begin{equation}\label{Jsum}
J_\mu(x) =\sum_\alpha J_\mu^{(\alpha)}(x)
\end{equation}
 and
\begin{equation}
J_\mu^{(\alpha)}(x)=\int  d^2\sigma \, \delta({\mathbf x}-{\mathbf y}^{(\alpha)}(\sigma))e^{a(\alpha)}_\mu(\sigma)\epsilon_{abc}\partial_b\xi_c^{(\alpha)}(x_0,\sigma)\, .
\end{equation}

Working in the Feynman gauge, the grounded partition function reads
\begin{equation}\label{ZgFeynman}
{\mathcal Z}^{(g)}= \int\Big[\prod_{\alpha=1}^N{\mathcal D}\xi^{(\alpha)}_a\Big] e^{-\frac{1}{2}\int_{x,x'}J_\mu(x)G(x-x')J_\mu(x')}\, ,
\end{equation}
with $G(x-x')$  the free Green's function of the four dimensional Laplacian. In terms of  its temporal Fourier transform 
\begin{equation}
\tilde G(\omega,{\mathbf x}-{\mathbf x'})=\frac{1}{4\pi}\frac{e^{-\vert\omega\vert\vert{\mathbf x}-{\mathbf x'}\vert}}{\vert{\mathbf x}-{\mathbf x'}\vert}\equiv
V_\omega({\mathbf x}-{\mathbf x'})\, ,
\end{equation}
we have

\begin{eqnarray}\label{jgj}
&&\int_{x,x'}J_\mu(x)G(x-x')J_\mu(x')\\ 
&& =\int_{{\mathbf{x}},{\mathbf{x'}}}\int\frac{d\omega}{2\pi}\tilde J^*_\mu(\omega,{\mathbf x})V_\omega({\mathbf x}-{\mathbf x'})\tilde J_\mu(\omega,{\mathbf x'})\equiv T^{(g)}\, .
\nonumber\end{eqnarray}

Now we implement the multipolar expansion. Generalizing the approach of Ref.\cite{Golestanian2000}, we will write Eq.(\ref{jgj}) in terms of the multipoles associated to the currents $J_\mu^{(\alpha)}$, that are linear functions of the auxiliary fields $\xi_a^{(\alpha)}$, and then perform
the functional integration in Eq.(\ref{ZgFeynman}) using as integration variables the multipoles instead of the auxiliary fields.

Combining Eqs.(\ref{Jsum}) and (\ref{jgj})  we see that $T^{(g)}$ can be written as a sum over all pairs of conductors
\begin{equation}
 T^{(g)}=\sum_{\alpha \beta}T^{(g)}_{\alpha\beta}\;,
 \end{equation}
 including self-pairings.

In the lowest order approximation, and when $\alpha\neq\beta$, we set:
\begin{equation}
V_\omega({\mathbf x}-{\mathbf x'})\simeq V_\omega({\mathbf x}_\alpha-{\mathbf x}'_\beta)\equiv V_\omega(R_{\alpha\beta})\, ,
\end{equation}
where ${\mathbf x}_\alpha$ denotes the position of the $\alpha_{ th}$ conductor and $R_{\alpha\beta}$ the distance between the pair of conductors.
Therefore
\begin{equation} \label{taneqb}
T^{(g)}_{\alpha\beta}\simeq \int\frac{d\omega}{2\pi}\tilde Q^{(\alpha)*}(\omega)V_\omega(R_{\alpha\beta})\tilde Q^{(\beta)}(\omega)\quad (\alpha\neq\beta)\, ,
\end{equation}
with
\begin{equation}
 \tilde Q^{(\alpha)}(\omega)=\int d^3 {\mathbf x}\,  \tilde J^{(\alpha)}_0(\omega,{\mathbf x})\, .
 \end{equation}
 
 Regarding  the diagonal terms $T^{(g)}_{\alpha\alpha}$, on general grounds, and neglecting higher order multipoles,  we expect to be proportional to square of the charge, that is
 \begin{equation}\label{taa}
 T^{(g)}_{\alpha\alpha} \simeq \int\frac{d\omega}{2\pi}\gamma_\alpha(\omega)\vert Q^{(\alpha)}(\omega)\vert^2
 \end{equation}
 for some frequency-dependent form factor $\gamma_\alpha$. We can obtain a crude estimation of  $\gamma_\alpha$ along the lines of
 Ref.\cite{Golestanian2000}.  We have:
 \begin{eqnarray}\label{taa2}
 T^{(g)}_{\alpha\alpha}&=& \int_{{\mathbf{x}},{\mathbf{x'}}}\int\frac{d\omega}{2\pi}\tilde J^{*(\alpha)}_0(\omega,{\mathbf x})V_\omega({\mathbf x}-{\mathbf x'})\tilde J_0^{(\alpha)}(\omega,{\mathbf x'}) \nonumber\\
 &=&\int\frac{d\omega}{2\pi} \int \frac{d^3{\mathbf{k}}}{(2\pi)^3}
 \frac{\vert\tilde{\tilde J}^{(\alpha)}_0(\omega,{\mathbf k})\vert^2}
 {{\mathbf{k}}^2+\omega^2}\, ,
\end{eqnarray}
where $\tilde{\tilde J}_0$  denotes the spatial Fourier transform of
${\tilde J_0}$. Let us assume that the characteristic size of the conductor
is $L_\alpha$.  The main contribution to the self-interaction comes 
from values of ${\mathbf {k}}$ such that $\vert{\mathbf {k}}\vert\leq 1/ L_\alpha$. Therefore, keeping only the monopole contribution in Eq.(\ref{taa2}) we obtain Eq.(\ref{taa}) with
\begin{equation}
\gamma_\alpha(\omega)=\int_{\vert{\mathbf {k}}\vert L_\alpha\leq 1}
\frac{d^3{\mathbf{k}}}{(2\pi)^3}
 \frac{1}
 {{\mathbf{k}}^2+\omega^2}\, .
 \end{equation}
 
We now insert the results for $T^{(g)}_{\alpha\beta}$ given in Eqs.(\ref{taneqb}) and (\ref{taa}) into Eq.(\ref{ZgFeynman}) and change the integration variables from the auxiliary fields $\xi^{(\alpha))}_a$ to the multipole moments. As the relation is linear the Jacobian is irrelevant for our purposes. 
In the monopole-monopole approximation we are keeping only the total charges and therefore
\begin{equation}
{\mathcal Z}^{(g)}\simeq \int\Big[\prod_{\alpha=1}^Nd\tilde Q^{(\alpha)}(\omega)\Big]\, e^{-\frac{1}{2}T^{(g)}}\, .
\end{equation} 
Performing the Gaussian integrals and introducing the frequency-dependent matrix:
\begin{equation}
\big[{\mathbb V}(\omega)\big]_{\alpha\beta}=\delta_{\alpha\beta}\gamma_\alpha(\omega)+(1-\delta_{\alpha\beta})V_\omega(R_{\alpha\beta})
\end{equation}
 we obtain
 \begin{equation}
 E^{(g)}\simeq\frac{1}{2}\int\frac{d\omega}{2\pi}\,{\rm tr}\log[{\mathbb V}(\omega)]\, .
 \end{equation}
 After standard manipulations we get, in the large distance limit
 \begin{eqnarray}
 E^{(g)}&\simeq &-\frac{1}{128\pi^2}\int d\omega \sum_{\alpha\neq\beta}\frac{1}{\gamma_\alpha(\omega)\gamma_\beta(\omega)}
 \frac{e^{-2\vert\omega\vert R_{\alpha\beta}}}{R_{\alpha\beta}^2}\\
 &=&-\frac{1}{128\pi^2}\sum_{\alpha\neq\beta}\frac{1}{R_{\alpha\beta}^3}\int d\nu \frac{e^{-2\vert\nu\vert}}{\gamma_\alpha(\nu/ R_{\alpha\beta})\gamma_\beta(\nu / R_{\alpha\beta})}
 \, , \nonumber 
 \end{eqnarray}
 where we discarded the self-energies of the conductors.
 The last equation shows that, for grounded conductors, the  vacuum energy decays as $1/R_{\alpha\beta}^3$ at large distances, as long as the form factors $\gamma_\alpha$
 have a well defined zero-frequency limit. This is indeed the case in the
 radial cutoff model used above, since the explicit form for
 $\gamma_\alpha$ becomes:
 \begin{equation}
	 \gamma_\alpha(\omega) \;=\; \frac{1}{2\pi^2 L_\alpha} \, [ 1 -
		 \omega L_\alpha \, {\rm arctan}(\frac{1}{\omega
		 L_\alpha})] \;,
 \end{equation}
 which tends to $\frac{1}{2\pi^2 L_\alpha}$ when $\omega L_\alpha  \to 0$.

 It is interesting to remark that higher order multipoles would contribute to both the grounded and isolated vacuum energies. However, the monopole terms are present only in the grounded case, because of the constraints on the total charges of the conductors. The leading contribution for isolated conductors would be the retarded dipole-dipole interaction, that produce a vacuum energy that decays as  $1/R_{\alpha\beta}^7$, as illustrated by the case of two spheres in the large separation limit \cite{Rodriguez-Lopez:2011spa}.  
  
\section{Conclusions}\label{sec:conc}

The main result of this paper is the observation that, when a system of conductors is connected to a charge reservoir, the vacuum forces receive additional contributions coming from charge fluctuations in each conductor. The difference between the vacuum energy for a system of isolated conductors and that corresponding to the same geometry but with grounded conductors depends on a  frequency-dependent capacitance 
matrix. 

In a multipolar expansion,   the monopole terms are responsible for the difference between the forces for isolated and grounded conductors. In particular, at large distances the vacuum energy decays as $1/R^3$ for grounded conductors, which corresponds to a retarded Coulombian interaction between fluctuating charges, while decays as $1/R^7$ in the isolated case, corresponding to the retarded dipole-dipole interaction (in the grounded case there is also a term that decays as $1/R^5$, due to the retarded monopole-dipole interaction).

We would like to conclude our paper with a discussion of the relation of our results with previous works on the subject. As already mentioned,
a similar difference between grounded and isolated conductors appears in the high temperature limit \cite{Golestanian2000, Fosco:2016rpu, BimonteEmig2012}, in which the problem reduces to an
electrostatic one. The nonretarded Casimir-Polder force on an atom in front of a perfectly conducting sphere also
depends on the grounding: the force decays as $1/R^4$ for a grounded sphere, and as $1/R^6$ for an isolated sphere \cite{Farina2010}. 

Charge induced fluctuation forces in graphitic nanostructures, similar to those considered here, have been described in Ref.\cite{Drosdoff}:
in that case, instead  of the presence of a charge reservoir,  charge fluctuations are allowed by connecting the conductors among them (the particular case considered there is that of  a parallel plate capacitor with the plates connected   by a wire). The charge density correlation is
evaluated using the fluctuation-dissipation theorem.

In the usual calculations of the Casimir effect  it is assumed (implicitly or explicitly) that the conductors are isolated. On the one hand, many works consider the case of perfect conductors as a particular limit of magneto-dielectric materials \cite{Teo:2012km}, which do not contain free charges. The perfect conductor limit then corresponds to isolated conductors.  When considering perfect conductors  in the canonical approach, a typical route is to choose the Coulomb gauge $\nabla\cdot{\mathbf A}$ and assume that, in addition, $A_0=0$ \cite{books}.  Note, however, that in the 
Coulomb gauge $A_0$
satisfies the Gauss law $\nabla^2 A_0=-\rho$. Although there are no charges in the volume between conductors, where $A_0$ satisfies the Laplace equation, fluctuating charges could be taken into account using nonvanishing  boundary conditions (that
should be properly averaged, as in
Ref.\cite{Drosdoff}).  Therefore, the usual assumption  $A_0=0$ imposes the neutrality condition. The energies and forces computed in this approach
are produced by the dipole and higher multipole fluctuations.

\section*{Acknowledgements}
This work was supported by ANPCyT, CONICET, UBA and UNCuyo.  We would like to thank C. Farina for pointing Ref.\cite{Farina2010} to us.

\end{document}